\documentclass[aps,prb,groupedaddress,twocolumn,floatfix,longbibliography,10pt]{revtex4-2}
\usepackage{amsmath}
\usepackage{amssymb}
\usepackage{graphicx}
\usepackage{xcolor}
\usepackage[inkscapelatex=false]{svg}
\usepackage{subcaption}
\usepackage[format=plain,justification=raggedright,singlelinecheck=false]{caption}
\usepackage{times}
\usepackage{hyperref} 
\hypersetup{colorlinks, allcolors=blue}

\newcommand{\mb}[1]{{\textcolor{black} {#1}}}

\begin{document}

\title{Spin-resolved quasiparticle interference patterns on altermagnets via non-spin-resolved scanning tunneling microscopy}
\author{Eric Petermann}
\email{eric.petermann@uni-wuerzburg.de}
\author{Kristian M{\ae}land}
\author{Bj{\"o}rn Trauzettel}
\affiliation{Institute for Theoretical Physics and Astrophysics, University of W{\"u}rzburg, D-97074 W{\"u}rzburg, Germany}
\affiliation{Würzburg-Dresden Cluster of Excellence ct.qmat, D-97074 W{\"u}rzburg, Germany}

\date{\today}

\begin{abstract}
We investigate quasiparticle interference on an altermagnetic Lieb-like lattice and show how a non-spin-polarized scanning tunneling microscopy measurement can yield effectively spin-resolved information. Within a four-site tight-binding model, which can be tuned between an antiferromagnetic and a Lieb-type altermagnetic state, we introduce on-site impurities at distinct sublattice sites and compute the real space local density of states (LDOS) via a Green's function approach. A Fourier transformation of the impurity-induced LDOS yields the characteristic $d$-wave spin-split Fermi surface contours of the altermagnetic phase. Notably, by choosing which sublattice the impurity is placed upon, we show that the scattering amplitudes effectively encode spin-dependent contrasts: Impurities on one of the magnetic sublattices highlights predominantly spin-up contributions along one crystallographic direction, while impurities on the other one favor the complementary spin-down channel and orientation.
\end{abstract}

\maketitle

\def\vec#1{\mathbf{#1}}

\section{Introduction}

Magnetism in solids is traditionally categorized into two basic collinear orders: ferromagnetism and antiferromagnetism. In a conventional ferromagnet, all magnetic moments align in parallel, yielding a net magnetization and spin split electron bands due to time reversal symmetry breaking. In contrast, a collinear antiferromagnet has antiparallel moments which cancel out, leading to a zero net magnetization and degenerate spin bands due to invariance under time reversal combined with translation or inversion. In the past years, a third distinct collinear magnetic phase called altermagnetism has been proposed \cite{Hayami2019Nov,Smejkal2020Jun,Yuan2020Jul,Ma2021May,Mazin2021Oct,Smejkal2022Sep,Smejkal2022Dec}. Altermagnets combine features of both ferro- and antiferromagnets. While they have a compensated net magnetization like an antiferromagnet, they exhibit significant momentum-dependent spin splitting of electronic bands like a ferromagnet. This nonrelativistic spin splitting arises from a broken antiferromagnetic translation symmetry while now preserving the system under a rotational symmetry combined with a time reversal operation. A wide range of material candidates like RuO${_2}$ \cite{Fedchenko2024Jan,Liu2024Oct,Smolyanyuk2024Apr}, CrSb \cite{Reimers2024Mar,Ding2024Nov,Yang2025Feb}, MnTe \cite{Mazin2023Mar,Lee2024Jan,Amin2024Dec}, MnF${_2}$ \cite{Bhowal2024Feb,Hariki2024Sep} and  Mn${_5}$Si${_3}$ \cite{Reichlova2024Jun,Rial2024Dec} have been suggested as potential altermagnets. 

A rather simple model for realizing altermagnetic order is the two-dimensional Lieb lattice \cite{Lieb1989Mar, brekkelieb, MaelandLieb24, franzlieblattice, Durrnagel2024AMLieb, Jakobinterfacelieb, Antonenko2025Lieb, Leraand2025Feb, Lundemo2025May, Syljuasen2025May, Jungwirth2024LiebRev, Takahashi2025May, Eskandari-asl2025Jul, Xu2025Lieb}. The Lieb lattice has a four-site unit cell in which only three sites are occupied by atoms, with the two diagonally opposed ones possessing a magnetic moment and the last one being non-magnetic. This lattice symmetry is invariant under a fourfold rotation followed by time reversal, but it breaks the usual translation symmetry that would interchange the two magnetic sublattices of a Néel antiferromagnet. As a result, the Lieb lattice can support an altermagnetic state. 

Finding a stable 2D electronic Lieb lattice has been challenging \cite{Slot2017LiebExp, Wu2024LiebExp}. The original motivation for introducing the Lieb lattice was to describe layers in high $T_c$ superconductors \cite{Lieb1989Mar}. Meanwhile, several recently proposed materials have Lieb-like layers and are altermagnetic candidates.
For example, the layered oxychalcogenide KV${_2}$Se${_2}$O crystallizes in a structure that within the layers realizes a 2D Lieb lattice of vanadium and oxygen atoms. It has been demonstrated to be a metallic, room-temperature altermagnet with a $d$-wave spin-momentum locking of its Fermi surface \cite{KVSEOSource}. The related compound Rb$_{1-\delta}$V${_2}$Te${_2}$O is another layered antiferromagnet now identified as a room-temperature altermagnet exhibiting $C$-paired spin-valley locking consistent with altermagnetic order \cite{RbVTeOsource}. Likewise, the Mn-based insulator La${_2}$O${_3}$Mn${_2}$Se${_2}$ consists of Mn$_2$O layers that form a Lieb lattice. It has been shown to host an antiparallel order with the expected altermagnetic band splitting \cite{LAOMNSESource, LAOMNSESource2}. \mb{See also Ref.~\cite{Chang2025Lieb} for a first-principle study providing further material candidates}. These examples underscore that Lieb-like sublattice geometries (and related structures) provide fertile ground for altermagnetism.

Altermagnets feature anisotropic, spin-split Fermi surfaces. Probing the Fermi surface (and its spin polarization) is therefore a key step toward identifying and characterizing altermagnetic order. Traditional angle-resolved photoemission can in principle detect spin-split bands, but it requires spin-sensitive detectors to distinguish the spin channels. An alternative approach is to use real space imaging of electronic standing waves via scanning tunneling microscopy (STM) and quasiparticle interference. In an STM experiment, the local density of states (LDOS) at the surface is measured via a tunneling current. Introducing a point impurity or defect on the surface causes electrons to scatter, producing spatial LDOS oscillations (Friedel oscillations) around the impurity. The patterns of these oscillations, known as quasiparticle interference (QPI) patterns, are directly related to allowed scattering vectors $\mathbf{q}$ connecting points on the constant-energy Fermi surface. 

By performing a Fourier transformation on the real space LDOS map, one obtains the FT-LDOS, which displays peaks at the $\mathbf{q}$-vectors corresponding to differences between Fermi momenta. In essence, the FT-LDOS is a momentum-space cut through the band structure at the chosen energy, revealing the Fermi surface contours. This Fourier-STM technique has been widely used to map out Fermi surfaces in a variety of quantum materials \cite{Crommie1993Jun,Avraham2018Oct}. For an altermagnet, one expects the QPI method to be especially informative, as it can visualize the distinctive spin-split Fermi surface features. However, because the spin-splitting in an altermagnet is present despite zero net magnetization, a conventional STM (with no spin polarization in the tip) will measure the combined LDOS of both spins. Hence, the discrimination of the two spin sub-bands seemingly requires spin-resolved STM techniques, which are experimentally challenging. 


Several theoretical works have recently addressed impurity-induced LDOS signatures in altermagnets. Based on continuum models, Refs.~\cite{jacob_impurity, Chen2024Friedel} make predictions about real space Friedel oscillations, while Refs.~\cite{Chen2024Friedel, Hu2025QPI} also show the FT-LDOS. On the momentum-space side, continuum model calculations of QPI have shown that impurities can indeed reveal the underlying spin-split Fermi surface. Their analyses indicate that spin-up and spin-down Fermi surface pockets give rise to distinct sets of scattering vectors. However, a spin-resoved STM measurement is required to distinguish the two spin split Fermi surfaces in terms of their spin index. The authors of Ref.~\cite{localsignatures} capture the lattice nature of real materials through a minimal two-sublattice model of altermagnetism. In real space, they found that the asymmetric part of the LDOS oscillations reflects the momentum-space nodal structure of the altermagnetic spin splitting. Remarkably, this signature is visible even in the spin-summed LDOS. These studies confirm that impurity-based STM can be a powerful tool to detect altermagnetic order. However, prior works have considered either continuum models or simplified two-sublattice models, without examining how a complex crystal geometry might affect the QPI. In particular, the inclusion of non-magnetic sites and the role of impurity location within a unit cell has received little attention so far.

In this work, we investigate quasiparticle interference on an altermagnetic Lieb-like lattice and demonstrate a strategy to obtain spin-resolved information from a non-spin-resolved STM measurement. We consider a four-site tight-binding model that can be tuned between a regular antiferromagnetic state and a Lieb lattice altermagnetic state by adjusting on-site potentials. By introducing a single impurity on different sublattice sites and computing the resulting LDOS patterns, we show that the choice of impurity sublattice effectively filters the spin content of the QPI. In the altermagnetic phase, impurities placed on one of the two magnetic sublattices produce QPI patterns dominated by states of a particular spin orientation (say, spin-up) along one crystallographic direction of the Fermi surface. Conversely, an impurity on the other magnetic sublattice highlights the opposite spin (spin-down) and the orthogonal Fermi surface direction. In this way, by comparing Fourier–STM images from impurities on different sublattices, one can reconstruct a spin-resolved map of the altermagnet’s Fermi surface – all using a standard (unpolarized) STM tip. This theoretical result suggests a practical experimental method to identify the spin-split Fermi surface of altermagnets without the need for specialized spin-sensitive instrumentation.

The article is organized as follows. In Sec.~\ref{sec:model}, we introduce the model with emphasis on the tuning of the magnetic phase achieved by changing on-site energies. In Sec.~\ref{sec:LDOS}, we focus on the real space electron densities of the system and how they are affected by placing impurities on different sublattice sites. In Sec.~\ref{sec:QPI}, we follow this up by examining the scattering images derived by Fourier transformation of the real space data. We conclude in Sec.~\ref{sec:Conclusions} and show additional plots and specify parameters in the Appendix.

\section{Model}
\label{sec:model}

\begin{figure*}[t]
    \centering
    \includegraphics[width=\linewidth]{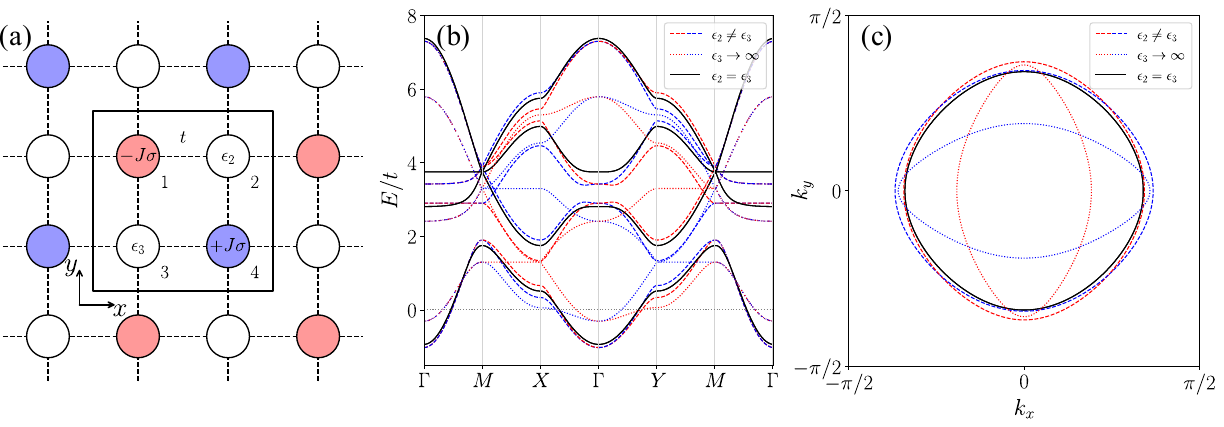}
    \caption{\textbf{(a)} Two-dimensional altermagnetic crystal lattice with two sites hosting opposite magnetic moments (up/red and down/blue) and two non-magnetic sites (white) in a checkerboard pattern. Itinerant electrons couple to the local moments through an exchange term $\pm J\sigma$, with $\sigma=\pm1$ for spin-up and spin-down carriers respectively. The values $\epsilon_{2/3}$ represent on-site energies for the non-magnetic sites and $t$ the hopping strength between nearest-neighbor bonds. By tuning $\epsilon_{2}$ and $\epsilon_3$, the system can be tuned between an antiferromagnetic (AFM) state with $\epsilon_{2}=\epsilon_{3}$ and an altermagnetic (AM) state with $\epsilon_{2}\neq\epsilon_{3}$. \textbf{(b)} \mb{ Band structure of the model along the high-symmetry-points for the parameters $(\epsilon_2=\epsilon_3=1)$ in black and $(\epsilon_2=0,\epsilon_3=1)$ and $(\epsilon_2=1,\epsilon_3\to\infty)$ in red (spin up) and blue (spin down).} In the case of $\epsilon_2\neq\epsilon_3$, the altermagnetic $d$-wave spin splitting becomes visible along the $\Gamma\to X$ and $\Gamma\to Y$ paths. \textbf{(c)} Fermi surface of the model.}
    \label{fig:latticeandbands}
\end{figure*}

We start with a simple four atom sublattice [Fig.~\ref{fig:latticeandbands}(a)] tight-binding model \cite{brekkelieb, MaelandLieb24, Jakobinterfacelieb} described by the Hamiltonian
\begin{align}
\nonumber
\mathcal{H} =& -t \sum_{\langle i,j \rangle,\sigma} c_{i,\sigma}^\dagger c_{j,\sigma}
- J \sum_{i,\sigma\sigma'} \mathbf{S}_i \cdot c_{i,\sigma}^\dagger \boldsymbol{\sigma}_{\sigma\sigma'} c_{i,\sigma'}\\ 
&- \sum_{i,\sigma} \left( \mu - \delta_{i,i_\text{NM}} \epsilon_\text{NM} \right) c_{i,\sigma}^\dagger c_{i,\sigma}.
\label{eq:hamiltonian}
\end{align}
Here, $c_{i,\sigma}^\dagger$ and $c_{i,\sigma}$ denote the creation and annihilation of an electron with spin $\sigma$ at lattice site $i$, respectively, while $t$ denotes the hopping strength between two nearest neighbors. 
A Kondo-like s-d coupling between the magnetic moment of the itinerant electrons and the magnetic moment of the local magnetic sites is added by the second term with a coupling strength $J$ in \mb{Eq.~\ref{eq:hamiltonian}}. 
\mb{$\boldsymbol{\sigma}$ is a vector of Pauli matrices, such that $\sum_{\sigma\sigma'}c_{i,\sigma}^\dagger \boldsymbol{\sigma}_{\sigma\sigma'} c_{i,\sigma'}$ represents the spin of the itinerant electron at site $i$. We separate the itinerant electrons and localized spins as two separate degrees of freedom and focus on N\'eel type ordering of the localized spins $\mathbf{S}_i$. Without loss of generality, we assume the localized spins order along the $z$-direction, for instance, due to easy-axis anisotropy. Then, the s-d coupling term becomes $- J \sum_{i,\sigma} \sigma S_i^z c_{i,\sigma}^\dagger c_{i,\sigma}$.}
We absorb the magnitude of $S_i^z$ into $J$ and assume site 1 has a local spin up and site 4 has local spin down. Sites 2 and 3 are nonmagnetic and
have on-site energies $\epsilon_\text{NM}\in(\epsilon_2,\epsilon_3)$. The Fermi level is controlled by $\mu$. 
\mb{We assume the effect of spin-orbit coupling on the electron band structure is negligible compared to the non-relativistic spin splitting, as expected in, e.g., KV${_2}$Se${_2}$O \cite{KVSEOSource}.}

In this model, the magnetic phases can be switched by tuning the on-site energies $\epsilon_\text{NM}$ of the non-magnetic sites. For the case $\epsilon_2=\epsilon_3$, the model possesses an antiferromagnetic order, as a translation between the two different magnetic sublattices and subsequent time-reversal operation leaves the system unchanged. Setting $\epsilon_2\neq\epsilon_3$ breaks this antiferromagnetic symmetry. Instead it is now invariant under fourfold rotation about the $z$-axis with a following time-reversal operation. In the limit of either $\epsilon_{2}\to\infty$ or $\epsilon_{3}\to\infty$, the lattice essentially becomes a Lieb lattice. The band structure of this model for the antiferromagnetic ($\epsilon_2=\epsilon_3$) and altermagnetic states ($\epsilon_2\neq\epsilon_3$) is illuminated in Fig.~\ref{fig:latticeandbands} (b). 
The figure shows the energy bands along high symmetry lines in the first Brillouin zone (1BZ), wherein we define the high symmetry points $\Gamma = (0,0)$, $X=(\pi/2,0)$, $Y = (0,\pi/2)$, and $M = (\pi/2, \pi/2)$.
Some of the recently proposed altermagnetic materials possess a surface structure that resembles a Lieb lattice \cite{KVSEOSource,LAOMNSESource,RbVTeOsource}. The key feature of the Lieb lattice is the vacancy in the unit cell. To have an altermagnetic order, a missing atom in the unit cell is however not necessary. One of the required components in creating the altermagnetic order in such a system is a lattice anisotropy between the non-magnetic sites.
\section{Local density of states}
\label{sec:LDOS}
\begin{figure*}[t]
    \centering
    \includegraphics[width=\linewidth]{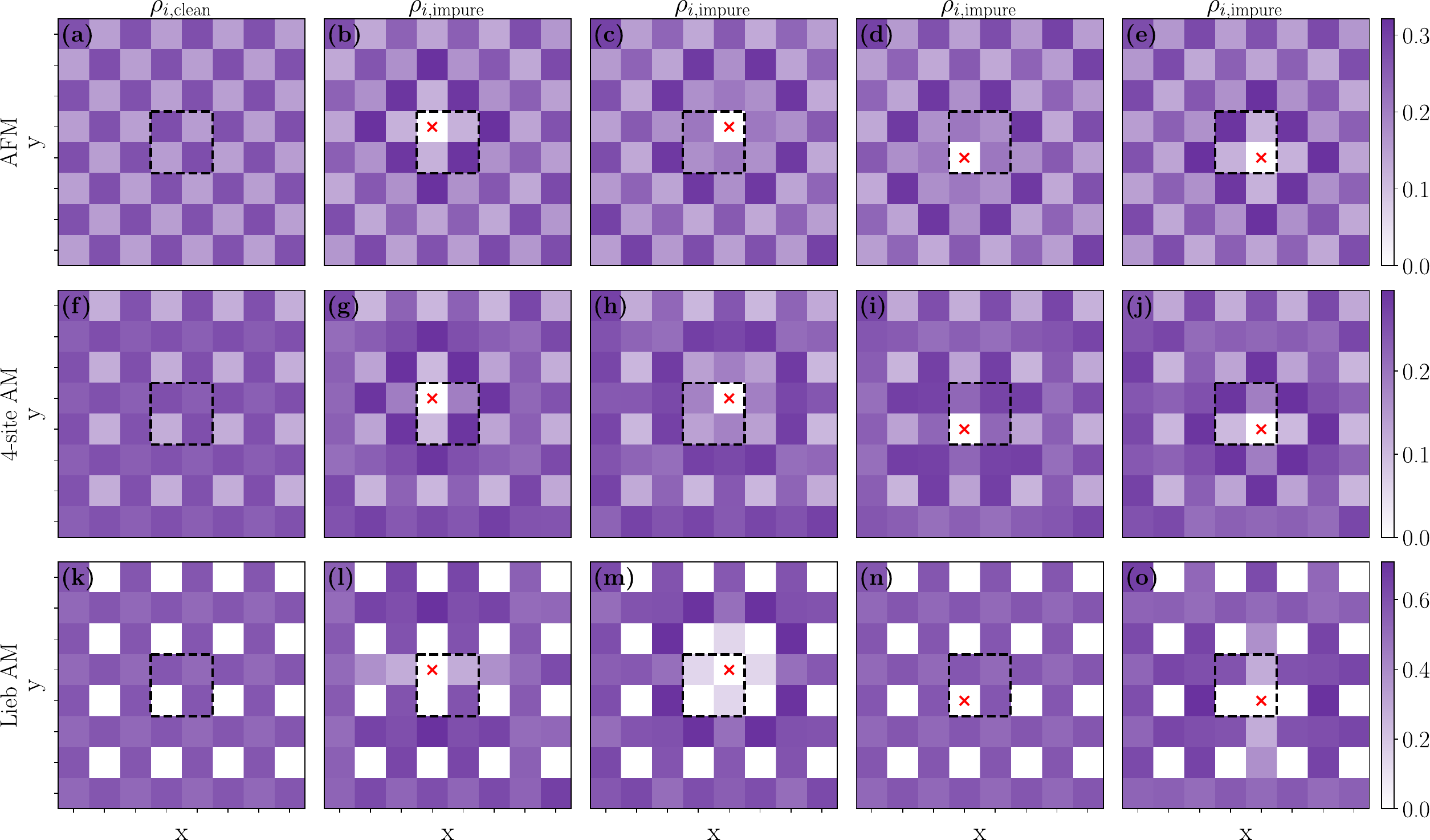}
    \caption{Real space Friedel oscillations in the charge LDOS for a single on-site impurity (red cross) placed on each of the four unit-cell sites (columns 2-5) with the lattice model tuned through three magnetic regimes (rows) at the Fermi level ($\omega=0$). Column 1 shows the undisturbed system. \textbf{(b-e)}: AFM state ($\epsilon_{2}=\epsilon_{3}$). \textbf{(g-j)}: 4-site AM state (finite $\epsilon_{2}\neq\epsilon_{3})$. \textbf{(l-o)}: Lieb lattice limit ($\epsilon_3\to\infty$).}
    \label{fig:LDOS_Sum}
\end{figure*}
\begin{figure*}[t]
    \centering
    \includegraphics[width=\linewidth]{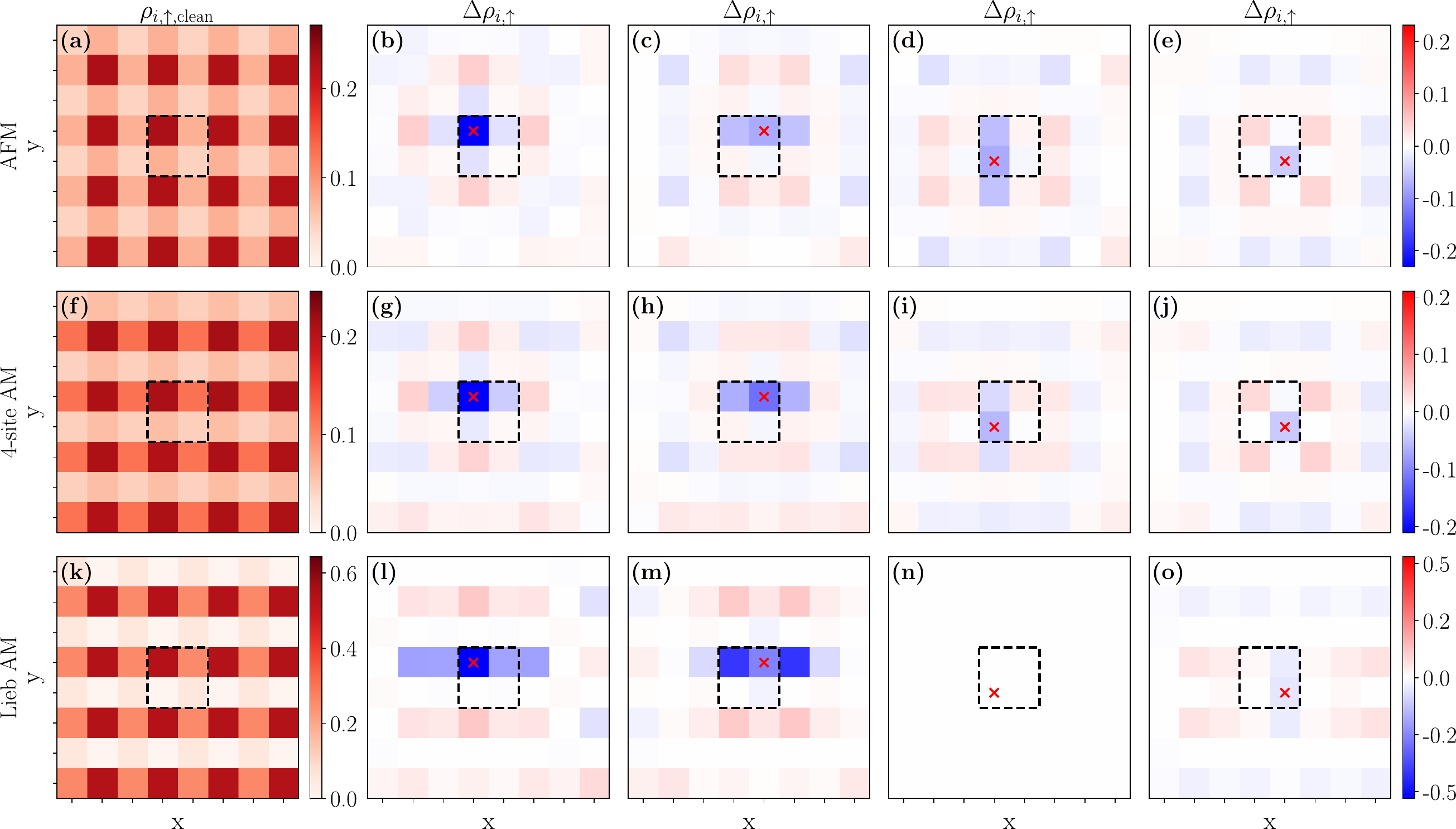}
    \caption{Clean spin-up LDOS and impurity-induced change in the spin-up LDOS, $\Delta\rho_{\uparrow} = \rho_{\uparrow,\text{impure}}-\rho_{\uparrow,\text{clean}}$, for an impurity (red cross) on each of the four unit-cell sites (columns) with the lattice model tuned through three magnetic regimes (rows) at the Fermi level ($\omega=0$). \textbf{(a-d)}: AFM state ($\epsilon_{2}=\epsilon_{3}$). \textbf{(e-h)}: 4-site AM state (finite $\epsilon_{2}\neq\epsilon_{3}$). \textbf{(i-l)}: Lieb lattice limit ($\epsilon_3\to\infty$). }
    \label{fig:LDOS_Up}
\end{figure*}
In this section, we analyze the real space local density of states (LDOS) to map out how electrons distribute across the lattice. For this purpose, we discretize our tight-binding Hamiltonian \eqref{eq:hamiltonian} on an $N\times N$ grid \mb{with hard-wall boundary conditions} and assemble it as a $2N^2\times2N^2$ matrix $H$, so that $\mathcal{H} =\Psi^\dagger H\Psi$ with
\begin{equation}
    \Psi =\bigl(c_{1,\uparrow},\dots,c_{N^2,\uparrow},\,c_{1,\downarrow},\dots,c_{N^2,\downarrow}\bigr)^T.
\end{equation}
We introduce a single impurity on one of the lattice sites by adding a non-magnetic impurity term given by $V=V_0\;c_{i,\sigma}^{\dagger} c_{i,\sigma}$ to the on-site energies of one of the central sites. We then calculate the Green's function of the system at a given energy $\omega$ as
\begin{equation}
    \label{eq:greens}
    G(\omega) = \left[ \omega + i\eta - H \right]^{-1},
\end{equation}
where $i\eta$ is a small imaginary shift. Mathematically, this places the poles of $G$ slightly below the real axis, making this a retarded Greens function enforcing causality. For numerical convenience, we keep the parameter $\eta$ finite ($\eta = 5\times 10^{-3}t$ throughout). A finite $\eta$ also mimics the finite energy resolution of an STM experiment. The choice of $\eta$ in the range $10^{-4}t\leq \eta \leq 10^{-1}t$ does not affect any conclusions of our work. 
The local density of states for spin $\sigma$ at a given lattice site $i$ then follows from
\begin{equation}
    \label{eq:LDOS}
    \rho_{i,\sigma}(\omega) = -\pi^{-1} \, \text{Im} \left[G_{ii,\sigma}(\omega) \right].
\end{equation}
The total electron occupation at each site is then $\rho_i(\omega)=\sum_\sigma \rho_{i,\sigma}(\omega)$.
The LDOS is calculated at the Fermi level ($\omega=0$) with $\mu$ chosen in a way that the Fermi level crosses the lowest two bands, regardless of how $\epsilon_1$ and $\epsilon_2$ may shift the bands. For the four-site and Lieb lattice altermagnetic cases we choose $\epsilon_3>\epsilon_2$. We also calculate the spin resolved effect of the impurity as $\Delta \rho_{i,\sigma} = \rho_{i,\sigma,\text{impure}}-\rho_{i,\sigma,\text{clean}}$ in order to explicitly show how the impurity affects the two spin flavors differently. We set the lattice constant to be $a=1$ in all further calculations following this section. All other system parameters are noted in Table~\ref{tab:parameters}. \mb{The LDOS can be directly measured in STM experiments, if the tip is carefully chosen such that its density of states is flat in the relevant energy range.}

Before discussing impurity effects, we first examine how the clean system LDOS depends on the relative on-site energies $\epsilon_2$ and $\epsilon_3$. The antiferromagnetic case ($\epsilon_2=\epsilon_3$), the altermagnetic case with four sites ($\epsilon_3>\epsilon_2$) and the Lieb lattice limit ($\epsilon_3\to\infty$) are shown spin-summed and for spin-up electrons in the first columns of Fig.~\ref{fig:LDOS_Sum} and Fig.~\ref{fig:LDOS_Up}, respectively. We choose to omit explicit spin-down images and refer to the symmetries connecting the two flavors when needed. 

In the antiferromagnetic case, we observe the expected checkerboard pattern in the spin-summed image, as the itinerant electrons accumulate on the favored magnetic sublattices. The spin-up image shows that the high occupancy on site 1 stems mostly from spin-up electrons, as this position is energetically favored by the on-site potential $-J\sigma$. For the same reason, site 4 exhibits a lower density, whereas sites 2 and 3 are populated equally. The resulting spin-up LDOS can be interpreted as having two quasi-one-dimensional LDOS channels passing trough site 1 horizontally along bonds 1-2 and vertically along 1-3. Conversely, spin-down electrons gather on site 4 with spin-down LDOS channels passing through it along bonds 2-4 and 3-4. 

Raising the on-site energy on sublattice 3 ($\epsilon_3>\epsilon_2$) switches the system into the altermagnetic regime. In the spin-summed image, this results in the loss of the checkerboard pattern. This is because occupancy of site 2 and 3 have now become unequal as a consequence of the different on-site energies. 
In the spin-up map we now observe a weakening of the vertical spin-up channel, as hopping along 1-3 bonds becomes less preferred. However, since both site 1 and 2 are still energetically favored in comparison to site 3 and 4, we have a strengthening of the horizontal spin channel. By the same argument, spin-down electrons now accumulate along the vertical 2-4 bonds, as the 3-4 
bonds have become more energetically unfavorable.

In the Lieb lattice limit ($\epsilon_3\to\infty$) site 3 becomes entirely unavailable for both spin flavors, as can be seen in the spin summed image. Consequently, the spin-up electrons now almost entirely congregate along the horizontal 1-2 bonds, while spin-down electrons almost entirely occupy the vertical 2-4 bonds channel.

We now place impurities on the central four sublattice sites. The resulting LDOS maps are shown non-spin-resolved in Fig.~\ref{fig:LDOS_Sum}~(Columns~2-5). For the spin-up LDOS, the impurity-induced change $\Delta \rho_\uparrow$ is shown in Fig.~\ref{fig:LDOS_Up}~(Columns~2-5). 
In the antiferromagnetic state, an impurity on site 1 produces isotropic scattering in the spin-summed LDOS, because site 2 and 3 have identical on-site energies. Most of the scattered electrons, however, are spin-up electrons, as both high-occupancy spin-up LDOS channels pass through site 1. Placing the impurity on site 4 reverses the spin roles. The spin-summed response remains isotropic, but is now dominated by spin-down electrons. For site 2 and 3 we also see directionally independent scattering in the summed image. The key difference is that in the spin-up image we observe the electrons mainly being scattered along the horizontal direction, as the 1-2 bonds possess a generally higher occupancy than the vertical 2-4 bonds. Following the same argument, spin-down electrons mainly scatter vertically with the same magnitude as spin-up electrons horizontally, creating the isotropic spin-summed scattering map. In the case of the impurity placement on site 3, the effect is reversed.

The choice of even a small difference between $\epsilon_2$ and $\epsilon_3$ compared to $t$ switches the system into an altermagnetic state. In the spin-summed LDOS, Friedel oscillations remain isotropic for impurities on sites 2 and 3, but become anisotropic for impurities on sites 1 and 4. Examining the spin-resolved maps clarifies why. Spin-up electrons scattered at the impurity on site 1 show a bigger deficiency along the horizontal direction compared to the vertical direction. This is because increasing $\epsilon_3$ has weakened the occupancy along the vertical spin-up LDOS channel while strengthening the horizontal channel. Hence, the impurity perturbs the more occupied bonds more strongly. Since the 1-2 and 1-3 bonds already host little spin-down electron weight in the clean system, their scattering response is weak in either direction. The imbalance between spin channels therefore produces an anisotropy in the spin-summed image. This effect is once again reversed on site 4, where the impurity instead has the highest impact on the vertical 2-4 bonds of the spin-down carriers. For impurities on site 2 and 3 the Friedel oscillations in the non-spin-resolved image remain isotropic. On site 2, each spin flavor has one channel whose LDOS increases with $\epsilon_3$ and one that decreases, but the preferred directions are opposite for up and down-spin electrons. Therefore, their sum is isotropic. On site 3, raising $\epsilon_3$ weakens the only remaining channel for each spin equally, making the impurity's impact comparable along both orthogonal directions and again preserving isotropy in the sum.

For the Lieb lattice limit, we choose $\epsilon_3\to\infty$, which is realized in the computational model by setting $\epsilon_3/t$ to a large number. This effectively makes the site unavailable for itinerant electrons of both spin flavors. In this limit, the previously described effects in the altermagnetic case become more pronounced. An impurity on site 1 produces strongly anisotropic scattering along the horizontal 1-2 bonds, dominated by spin-up electrons. Conversely, an impurity on site 4 yields almost exclusively spin-down electron scattering along the vertical 2-4 bonds. For site 2, we find mainly horizontal depletion in the spin-up channel and a vertical response in the spin-down channel, which again sums to an isotropic LDOS. Finally, placing the impurity on site 3 has essentially no effect on either spin, since the site is already energetically inaccessible.

\section{Quasiparticle interference pattern}
\label{sec:QPI}
\begin{figure*}[t]
    \centering
    \includegraphics[width=\linewidth]{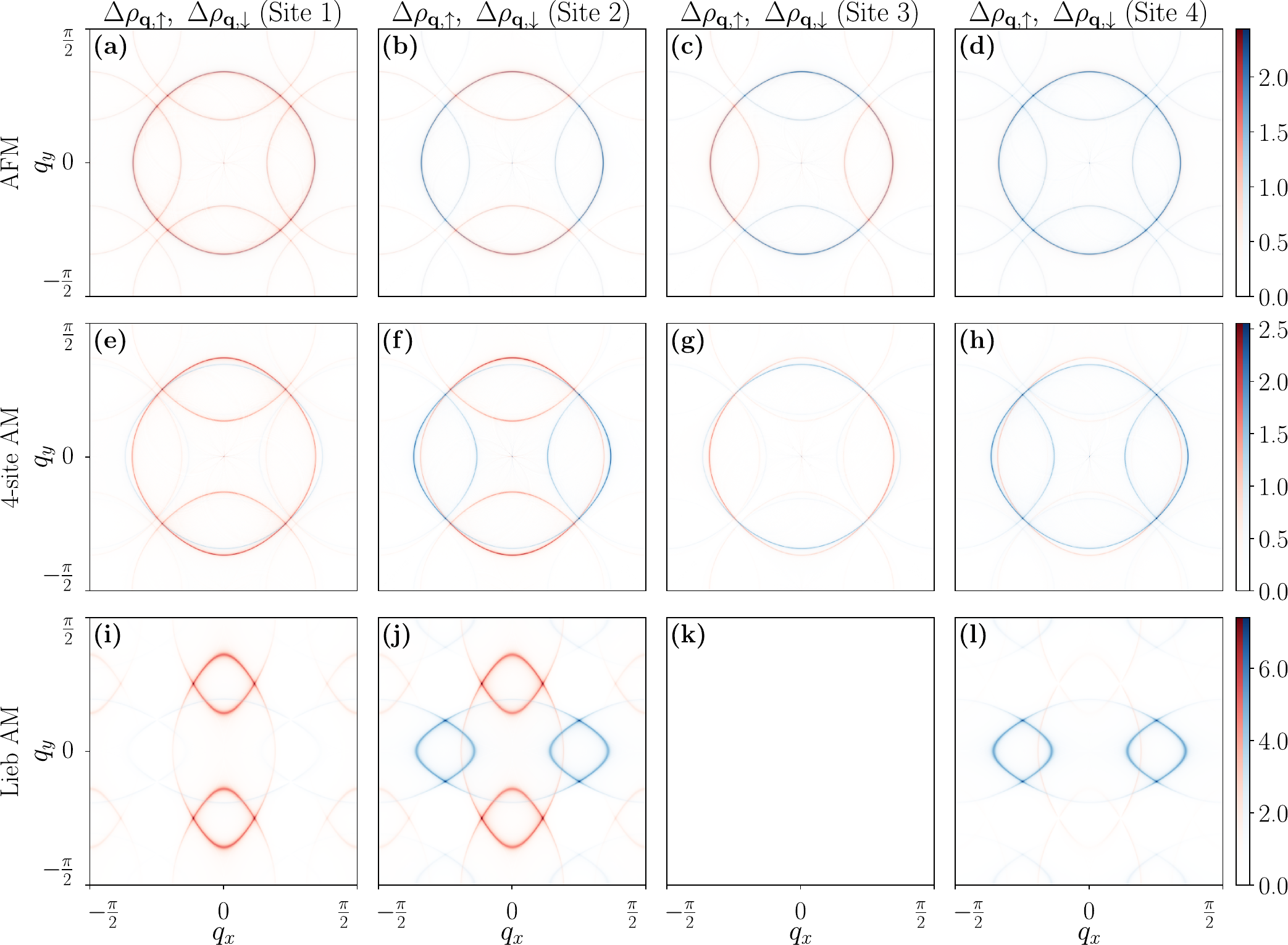}
    \caption{Absolute value of the Fourier-transformed LDOS (power spectrum) at the Fermi level ($\omega=0$) for a single impurity placed on each of the four unit-cell sites (columns) with the lattice model tuned through three magnetic regimes (rows). The intensity gives the spin summed power spectrum, while red and blue denote spin up and spin down contributions, respectively. \textbf{(a-d)}: AFM state ($\epsilon_{2}=\epsilon_{3}$). \textbf{(e-h)}: 4-site AM state (finite $\epsilon_2\neq\epsilon_3$). \textbf{(i-l)}: Lieb lattice limit ($\epsilon_3\to\infty$).}
    \label{fig:FTLDOS_Sum}
\end{figure*}

Let us now move from real space to momentum space. By choosing a large enough system size, the Fourier transformed local density of states (FT-LDOS) is then calculated using a fast Fourier transformation (FFT) at the Fermi level. Before conducting the FFT the system without the impurity is subtracted from the impure system in order to remove self-correlation peaks from the resulting FT-LDOS image. The FT-LDOS is
\begin{equation}
    \Delta\rho_{\mathbf{q},\sigma} = \sum_i \Delta\rho_{i,\sigma}e^{-i\mathbf{q}\cdot\mathbf{r}_i},
\end{equation}
where $\mathbf{r}_i$ is the location of site $i$ and $\mathbf{q}$ is a quasimomentum restricted to the 1BZ.
We take the absolute value of this transformation to get the power spectrum $|\Delta\rho_{\mathbf{q},\sigma}|$. The factor $\mu$ is once again tuned in a way that the Fermi level crosses the lowest two bands (the parameters are the same as for the LDOS calculations). This yields a visualization of the scattering vectors ${\mathbf{q}=\mathbf{k}'-\mathbf{k}}$ on a constant energy slice. The FT-LDOS has been shown to reproduce the Fermi surface of a low energy effective model of an altermagnet \cite{jacob_impurity,Hu2025QPI,Chen2024Friedel}. We plot the spin-resolved FT-LDOS in Fig.~\ref{fig:FTLDOS_Sum}. 
These panels are shown spin-resolved for clarity, but if the color coding were removed the image would reduce to the spin-summed FT-LDOS, yielding the same scattering pattern and strengths. However, as we argue below, spin-resolved information can be obtained by non-spin-resolved measurements of the QPI pattern.

In the antiferromagnetic state ($\epsilon_2=\epsilon_3$), the non spin-resolved FT-LDOS produces the same isotropic ($s$-wave) scattering contour, so the two spin channels cannot be separated by intensity alone. The spin-resolved maps, however, reveal that scattering around site 1 is dominated by spin-up electrons, whereas scattering around site 4 is dominated by spin-down electrons. This matches with the explanation in Sec.~\ref{sec:LDOS}, with high-occupancy LDOS channels being more affected by the impurity. The same holds true for site 2 and 3, where although the two spin-bands are degenerate, we see directionally dependent scattering strengths. 
The FT-LDOS also displays a periodic repetition of the Fermi surface due to scattering to neighboring Brillouin zones.
This feature is not captured by low-energy effective models. Finally, note that all Fermi-surface contours in the FT-LDOS are enlarged by a factor of two. This is because, for example, scattering from Fermi vector $-k_F$ to $k_F$ leads to an effective scattering vector of $|\mathbf{q}|=2k_F$. 

Switching to $\epsilon_2\neq\epsilon_3$, the $d$-wave altermagnetic dispersion clearly becomes visible in all of the FT-LDOS images. For an impurity on site 1 this deformation is accompanied by a pronounced enhancement of the spin-up channel and a simultaneous suppression of the spin-down channel. Site 4 shows the opposite behavior, exactly as anticipated from the real-space analysis in Sec.~\ref{sec:LDOS}. On site 2 the overall scattering intensity grows for both spins, as the horizontal and vertical LDOS channels of the respective spins that cross the site gain spectral weight as $\epsilon_3$ increases. Hence, they are more strongly perturbed by the impurity. By contrast, an impurity on site 3 now produces a weaker response, as the site is already less occupied because of the higher on-site energy. Most importantly, the anisotropic $d$-wave contours now lets us distinguish the two spin channels on sites 1 and 4 by their scattering amplitudes alone.

Lastly, we move to the Lieb lattice limit with $\epsilon_3\to\infty$. The $d$-wave altermagnetic shape becomes even more pronounced in the scattering image. On site 2, we clearly see the full $d$-wave altermagnetic Fermi surface with similar scattering strength along $x$ and $y$ direction. 
The two-sublattice model of Ref.~\cite{localsignatures} demonstrates that a continuum model is not sufficient to fully capture the QPI around an impurity in altermagnets. Interestingly, by including additional nonmagnetic sublattices, we find that if the impurity is placed on a nonmagnetic site, similar patterns as predicted in the continuum models \cite{jacob_impurity, Hu2025QPI, Chen2024Friedel} appear. Namely, the intensity of the spin up and spin down Fermi surfaces are the same.
In the Lieb lattice case, an impurity on site 3 produces almost no signal, as the additional potential barely perturbs a site that is already energetically inaccessible. In contrast, impurities on sites 1 and 4 yield spectra dominated by spin-up electrons and spin-down electrons, respectively. Since the scattering amplitude on each of these sites is strongly spin-polarized, we can extract an effectively spin-resolved image from the spin-summed FT-LDOS simply by knowing which magnetic sublattice hosts the impurity. 

The Lieb lattice case also highlights the anisotropy in the intensity of the power spectrum along the Fermi surface. For an impurity on site 1, the intensity is strongest for large $|q_y|$ and small $|q_x|$, see Fig.~\ref{fig:FTLDOS_Sum}(i). This can be understood from the real space LDOS in Fig.~\ref{fig:LDOS_Up}(l) showing slow, long-wavelength Friedel oscillations in the $x$ direction (small $|q_x|$) and rapid, short wavelength oscillations in the $y$ direction (large $|q_y|$). For an impurity on site 4, a similar anisotropic power spectrum emerges for similar reasons.
We again emphasize that this effect depends on the presence of the non-magnetic sublattices. 

\mb{A potential challenge in experiments is how to determine which magnetic sublattice the impurity is placed on. STM topography measurements can be used to find the location of the impurity, and assuming a Lieb like lattice, the two magnetic sites can be identified without any spin-resolved STM measurement (assuming the magnetic order has been confirmed by other means). Then, one does not need to identify which sublattice is spin up and which is spin down, simply that two measurements done on the same geometry have the impurity placed on two different magnetic sublattices. The predicted difference in the QPI patterns for the two cases can then be checked, and the results can be interpreted as the two spin-split Fermi surfaces, without necessarily knowing which is spin up and which is spin down. We also perform calculations with several impurities placed on the same sublattice and find similar QPI patterns.}

\section{Conclusions}
\label{sec:Conclusions}
Placing impurities on the surface of certain altermagnetic candidates (which can be mapped onto a Lieb lattice) leads to Friedel oscillations in the local density of states. By Fourier transformation, the scattering vectors of the system can be derived. Plotting the scattering vectors show the altermagnetic Fermi surface. Notably, it can be seen that placing the impurity on the magnetic sites specifically yields additional information about the spin flavor. Depending on which magnetic site the impurity is placed on, the spin is encoded in the scattering magnitudes of the electron wave functions in the LDOS. Even though no spin-resolved measurement is assumed, effectively spin-resolved images of the Fermi surface can be obtained. Hence, this proposal presents an alternative way to obtain information of the spin-split bands of altermagnets by standard (spin-unresolved) STM imaging.
\begin{acknowledgments}
We thank Changan Li, Johannes Mitscherling, Artem Odobesko, and Pavlo Sukhachov for useful discussions. This work was supported by the Deutsche Forschungsgemeinschaft (DFG, German Research Foundation) project SFB 1170 and DFG through the Würzburg-Dresden Cluster of Excellence ct.qmat (EXC 2147, project-id 390858490).
\end{acknowledgments}

\appendix
\section{Numerics}
\label{app:Numerics}
All numerical calculations are carried out on a $502\times502$ square lattice comprising $251\times251$ unit cells with open boundary conditions. This chosen system size ensures that the single on-site impurity, placed exactly at the central unit cell, is separated by at least $250a$ from any edge. This way, the induced Friedel oscillations decay well before reaching the boundary. Additionally, choosing a larger system yields a better resolution in the FFT image of the grid. The Hamiltonian \eqref{eq:hamiltonian} is assembled as a sparse matrix. The inversion required for the Green's function \eqref{eq:greens} is conducted using a block-by-block inversion algorithm designed for tridiagonal block matrices \cite{Reuter2012Jul}. To keep the matrix tridiagonal, non-periodic boundaries are also required. We set $\eta=5\times10^{-3}$. A non-zero $\eta$ smooths the discrete spectrum one would otherwise receive by exact diagonalization, so that the Fourier transformed density of states in Sec.~\ref{sec:QPI} captures the spectrum accurately. Tests with $\eta$ in the range of $10^{-4}t\leq\eta\leq10^{-1}t$ yield similar results. All system parameters are shown in Table~\ref{tab:parameters}.
  
\begin{table}[ht]
  \centering
  \caption{Table showing the parameters for the three different magnetic states. In all cases we set $J=t=1$, the impurity potential $V_0=100$, and the lattice constant $a=1$.}
  \begin{tabular}{cccc}
    \hline\hline
    & $\epsilon_2$ & $\epsilon_3$ & $\mu$\\ \hline
    AFM         & $1$        & $1$         & $-2.75$         \\ 
    4-site AM         & $0$         & $1$         & $-2.90$         \\ 
    Lieb AM         & $1$        & $10^7$         & $-2.30$         \\ \hline\hline
  \end{tabular}
  
  \label{tab:parameters}
\end{table}

\bibliography{library}

\end{document}